# Low-energy Electron Reflectivity from Graphene: First-Principles Computations and Approximate Models

R. M. Feenstra and M. Widom
Dept. Physics, Carnegie Mellon University, Pittsburgh, Pennsylvania 15213, USA

**Abstract**

A computational method is developed whereby the reflectivity of low-energy electrons from a surface can be obtained from a first-principles solution of the electronic structure of the system. The method is applied to multilayer graphene. Two bands of reflectivity minima are found, one at 0 – 8 eV and the other at 14 – 22 eV above the vacuum level. For a free-standing slab with $n$ layers of graphene, each band contains $n-1$ zeroes in the reflectivity. Two additional image-potential type states form at the ends of the graphene slab, with energies just below the vacuum level, hence producing a total of $2n$ states. A tight-binding model is developed, with basis functions localized in the spaces between the graphene planes (and at the ends of the slab). The spectrum of states produced by the tight-binding model is found to be in good agreement with the zeros of reflectivity (i.e. transmission resonances) of the first-principles results.

## I. Introduction

Over the past several years, many researchers have used the low-energy electron microscope (LEEM) to measure the reflectivity vs. electron energy of electrons from single- and multi-layer graphene. As first demonstrated by Hibino and co-workers in 2008, such spectra reveal characteristic oscillations over the energy range 0 – 8 eV, with the number of reflectivity minima being directly related to the number of graphene layers on the surface [1]. Hence, reflectivity spectra acquired pixel-by-pixel over some field of view (i.e. by acquiring a sequence of images at different energies) permit detailed mapping of the graphene thickness. Such measurements have been pursued for graphene on $SiO_2$ [2], on SiC [3,4,5] and on various metal substrates [6,7,8]. In two recent papers, we have argued that the oscillations in the reflectivity spectra from graphene permit not only the determination of the thickness of the multilayer graphene, but they also can yield information about the interface between the graphene and the substrate that it resides on [9,10]. Such measurements are also expected to be useful for other two-dimensional (2D) materials as well.

The oscillations in the reflectivity originate from *interlayer states* that occur in multilayer graphene and other 2D materials [9,11]. These states arise from the exchange-correlation potential of the system in the same way that image-potential states at the surface of crystals are formed: when an electron is removed from a surface (or introduced near a surface) a region of positive charge, a so-called correlation hole, is formed in the material. An attractive interaction is formed between the electron and the hole, with asymptotic form of $1/4|z|$ where $z$ is the separation between the surface and the electron [12]. (The factor of 4 here arises from two separate factors of 2: one of these occurs simply because the distance between the electron and the correlation hole is $2|z|$, and the other occurs since the vacuum exists only over a half-space [13]). This $1/4|z|$ potential produces a bound state for the electron at the surface (an entire series of Coulombic-like states in fact [12]) and the ground state of this series we refer to as the image-potential state of the surface.

For the case of 2D materials, the separation between atomic planes is sufficiently large so that states similar to the image-potential state of the surface can form. For a thick slab of 2D material, combinations of these *interlayer states* form a band, the so-called *interlayer band*, first identified and characterized by Posternak et al. [11]. The energy of the interlayer states/band is higher than that of the image-potential surface states, due primarily to confinement of the interlayer states. In graphene, these states occur in the energy range 0 – 8 eV, where they are ideally situated for observation using a LEEM.

In our prior work we developed a method for deriving the electron reflectivity from first-principles computations of the electronic structure of the material [9]. Our method is similar to that of Krasovskii and co-workers [14], except that: (i) their computations are for the surface of a semi-infinite system whereas our computations use a periodically repeated slab structure in the *z*-direction (perpendicular to the surface); and (ii) unlike Krasovskii et al. we do not incorporate an imaginary term in the potential, and hence the fall-off in electron reflectivity at higher energies (due to inelastic effects) is not present in our theory. Additionally, the method of Krasovskii is further developed than our own, having been applied to many more situations. Nevertheless, our computational method permits a straightforward evaluation of reflectivity using a wide-spread electronic structure code, the Vienna Ab-Initio Simulation Package (VASP) [15,16]. Therefore, it perhaps can be more easily applied by a wider community of users, e.g., those workers already familiar with VASP or similar computer codes. In any case, the results of the two methods should be identical, aside from the aforementioned lack of imaginary term in the potential for our computations.

In addition to first-principles computations, there are approximate models that have been utilized in the past for providing some understanding of reflectivity spectra. For graphene in particular, Hibino et al. proposed a tight-binding type of model in which minima of the reflectivity (*transmission resonances*) were understood in terms of discrete states in the tight-binding scenario.[1] Some modification of this model has been proposed in our prior work, replacing states derived from the atomic planes with ones existing in the spaces between the planes, and introducing states with different energy in the space between the substrate and the nearest graphene plane [9,10]. We further discuss this tight-binding model in the present work, providing an explicit set of matrix elements that yields a spectrum of states similar to those obtained by first-principles.

Other types of models that have been used in the past for describing reflectivity spectra include a one-dimensional (1D) solution of Schrodinger's equation using some approximate potential [17], or the phase-accumulation (PA) model that is generally applied to metallic films for which the potential in the material can be reasonably treated as a constant (i.e. the *inner potential* of the material). The latter model has been successfully employed for describing quantum-well type states that occur in the thin metal films [18,19]. We provide a brief discussion here concerning the applicability of these models for multilayer graphene.

In the following Section we summarize the first-principles method, providing plots of wavefunctions for typical cases in the analysis, and also describing the full range of wavefunctions that occur, including surface image-potential states and two bands of interlayer states (the first at 0 – 8 eV and the second at 14 – 22 eV). In Section III we describe a tight-binding model that can be used to reproduce the reflectivity minima found in the first-principles computations. In Section IV we discuss other models for describing the reflectivity of graphene, and in Section V the paper is summarized.

## II. First-principles Computations

Let us first consider a free-standing graphene slab. Computations are performed in a periodically repeated vacuum-graphene-vacuum system, with at least 1 nm of vacuum separating the graphene slabs. The geometry is pictured in Figs. 1(a) and 1(b), for the case of a 4-layer graphene slab. The direction perpendicular to the surface of the slab is labeled by $z$, and the directions parallel to the slab are labeled by $x$ and $y$. A view of the local potential (ionic plus Hartree plus exchange-correlation) for this system, as obtained from VASP and then averaged over the $xy$ plane, is shown in Fig. 1(c). The location of the *vacuum level* corresponds to the potential far from the graphene slab; we use this to define our zero of energy. It should be noted that VASP employs the projector-augmented wave (PAW) method [17] (similar to the pseudopotential method) which smooths over the diverging potential that would occur very near the nucleus of each atom. The PAW-wavefunctions are, outside the atomic cores, identical to those that would be obtained in an all-electron solution of the Kohn-Sham equations, but near the cores they satisfy a different set of conditions. The energy eigenvalues are, in principle, identical for the PAW method and an all-electron method. We use the Perdew-Burke-Ernzerhof (PBE) generalized gradient approximation to the exchange correlation functional [20].

For an incoming electron beam at normal incidence, we need only consider states in the graphene having Bloch-wavevector **k** with components $k_x = k_y = 0$ and $k_z \neq 0$, although these states do, crucially, contain components with lateral wavevector $(G_x, G_y)$, where $\mathbf{G} \equiv (G_x, G_y, G_z)$ is a reciprocal lattice vector. Figure 2 shows the energy eigenvalues for the 4-layer graphene slab plotted with respect to $k_z$, for states located 0 – 10 eV above the vacuum level. A dispersive band is seen to be present in this range; its origin is identical as that for the *interlayer band* of bulk graphite [11]. The states within this band are referred to as *interlayer states*, having character as described previously by Posternak et al. with wavefunctions that are concentrated in the spaces between the graphene planes [11]. The large number of sub-bands and the limited range of wavevectors in Fig. 2 is a result of the four layers of graphene (16 atoms) and the large amount of vacuum (here, the periodicity in the $z$-direction is 4.0266 nm).

In order to compute the reflectivity, it is necessary to match planes waves existing in the vacuum region to states within the graphene slab. For this purpose it is important to note that the states in the slab have a quite distinct character: classically, we can have electrons moving in the vacuum perpendicular to the surface (which is what we desire for the matching) or not parallel to it (which would correspond to diffracted beams). The latter states require a sufficiently large perpendicular component in their wavevector to overcome the potential barrier at the surface of the slab energy and become freely propagating in the vacuum. That is, for a lateral wavevector of $(G_x, G_y)$ such states require an energy, relative to the vacuum level, greater than $\hbar^2 (G_x^2 + G_y^2)/2m$. For energies less than this they are confined in the slab as evanescent states, which decay into the vacuum.

On Fig. 2 we mark several typical states in the energy bands, two at energies of 7.09 and 7.14 eV associated with flat bands (shown more clearly in the inset on the right-hand side), and one at energy near 3.24 eV associated with the dispersive band. Wavefunctions for these states are shown in Fig. 3. States (a) and (b) form a pair, with odd and even symmetry, respectively, relative to $z = 0$. These states decay nearly to zero as the distance away from the graphene slab increases; they correspond predominantly to electrons moving laterally in the graphene slab (with wavevector equal to a lateral reciprocal lattice vector),[1] and they correspondingly display oscillatory character in the $x$ and $y$ directions as seen on the far right-hand side of the plot. These states will *not* couple to plane waves in the vacuum, and hence we discard them in our analysis method. The precise numerical test by which we distinguish these states is defined elsewhere [9].

The state of Fig. 3(c) displays oscillatory character in the vacuum on either side of the slab, and nearly no lateral variation. This state *will* couple to plane waves in the vacuum. At a large enough distance from the graphene slab such that the potential is nearly constant (≥0.5 nm), then the wavefunction for this state reduces to simply a linear combination of incoming and outgoing plane waves, i.e. with no dependence on the $x$ or $y$ coordinates. For matching to these plane waves in the vacuum, it is convenient to work with the Fourier components of the wavefunction. States are labeled by $\psi_{\nu,k_z}(\mathbf{r})$ with energy $E_{\nu,k_z}$, where $\nu$ is a band index. A state can be decomposed according to

$$\psi_{\nu,\mathbf{k}}(\mathbf{r}) = \sum_{\{\mathbf{G}\}} \frac{C_{\nu,\mathbf{G}}}{\sqrt{V}} e^{i(\mathbf{k}+\mathbf{G})\cdot\mathbf{r}} = \sum_{\{G_x,G_y\}} \phi_{\nu,k_z}^{G_x,G_y}(z) e^{i[(k_x+G_x)x+(k_y+G_y)y]} \qquad (1)$$

where $C_{\nu,\mathbf{G}}$ is a plane-wave expansion coefficient for reciprocal lattice vector $\mathbf{G}$, $V$ is the volume of a unit cell (included for normalization purposes), and with

$$\phi_{\nu,k_z}^{G_x,G_y}(z) = \sum_{\{G_z\}} \frac{C_{\nu,\mathbf{G}}}{\sqrt{V}} e^{i(k_z+G_z)z}. \qquad (2)$$

Our evaluations are performed for $\mathbf{k} = (0,0,k_z)$, so we need only consider $\psi_{\nu,k_z}(\mathbf{r})$ in comparison to $\phi_{\nu,k_z}^{G_x,G_y}(z)$. For the state pictured in Fig. 3(c), far out in the vacuum its total wavefunction has purely $\phi_{\nu,k_z}^{0,0}(z)$ character, i.e., for states such as this, $\psi_{\nu,k_z}(\mathbf{r}) \to \phi_{\nu,k_z}^{0,0}(z)$ for $|z| \to \infty$. Our analysis could proceed entirely in terms of $\psi_{\nu,k_z}(\mathbf{r})$, but it is more convenient to use $\phi_{\nu,k_z}^{0,0}(z)$ instead since this function has no *x* or *y* dependence whatsoever, whereas $\psi_{\nu,k_z}(\mathbf{r})$ retains some slight *x,y* dependence in the vacuum region.

---

[1] More precisely, since we are considering $k_x = k_y = 0$, the states of Figs. 3(a) and 3(b) have parallel wavevectors that are equal to reciprocal lattice vector of the slab, and hence these states are standing wave states formed at the edge of a Brillioun zone. In the reduced-zone scheme, this lateral wavevector is then folded into the zone center.

The procedure that we use to deduce the reflectivity for a free-standing graphene slab is described in Ref. [9]. We evaluate both $\phi_{\nu,k_z}^{0,0}(z)$ and $\phi_{\nu,-k_z}^{0,0}(z)$, and from these two wavefunctions we then construct a state with only outgoing plane wave character on the right-hand side of the slab and with both incoming and outgoing character on the left. The first step in this procedure is to form both symmetric and antisymmetric combinations of $\phi_{\nu,k_z}^{0,0}(z)$ and $\phi_{\nu,-k_z}^{0,0}(z)$, namely, $\phi_{\nu,\pm}^{0,0}(z) \equiv [\phi_{\nu,k_z}^{0,0}(z) \pm \phi_{\nu,-k_z}^{0,0}(z)]/\sqrt{2}$. These states have standing wave character, i.e. with real and imaginary parts that are equal to each other within a proportionality factor; for convenience, they are normalized such that the resulting functions are real. It is useful to characterize these standing wave states in terms of phase shifts, expressing them, respectively, in the vacuum region as $A_+ \cos(\kappa_0 z \pm \delta_+)$ and $A_- \sin(\kappa_0 z \pm \delta_-)$, with the upper sign referring to the left-hand side of the vacuum region and the lower sign the right (actually, these forms are valid for a potential that is symmetric with respect to *z*; the more general case is treated in Ref. [9]). The phase shifts $\delta_+$ and $\delta_-$ can be obtained simply by locating the zeroes in $\phi_{\nu,+}^{0,0}(z)$ and $\phi_{\nu,-}^{0,0}(z)$. Given these simple sine and cosine type wavefunctions, it is straightforward to further construct a state that has only outgoing, plane wave type character on the right-hand side of the slab. In doing so, a solution for the reflectivity is obtained, which for a symmetric potential turns out to be

$$R = \sin^2(\delta_+ - \delta_-). \tag{3}$$

The corresponding formula for a general potential is provided in Ref. [9].

Figure 4 shows results for the reflectivity of a 4-layer graphene slab. The reflectivity is plotted over the range 0 – 22 eV, showing two bands located at 0 – 8 eV and 14 – 22 eV where oscillations in the reflectivity occur. The location of two image-potential surface states at $-0.59$ and $-0.68$ eV relative to the vacuum level are indicated in the reflectivity plot of Fig. 4. Because they lie below the vacuum level, the states are not observed in LEEM reflectivity. Wavefunctions for those two states are displayed in Figs. 4(a) and 4(b). Figures 4(c) – 4(e) show wavefunctions at the energies corresponding to the minima (zeroes) in the reflectivity for the first band (0 – 8 eV), and Figs. 4(f) – 4(h) show wavefunctions for the second band (14 – 22 eV). For these states 4(c) – 4 (h), it should be noted that the wavefunctions in the vacuum are essentially constant, i.e. corresponding simply to an incoming plane wave from the left and an outgoing plane wave on the right-hand side of the graphene slab. This result is in contrast to that obtained for all other states with specific $k_z$, e.g. Fig. 3(c), for which an oscillatory nature for the wavefunction magnitude in the vacuum is obtained. Our general method for the reflectivity analysis, summarized above, combines states at $+k_z$ and $-k_z$ such that the wavefunction magnitude on the right-hand side of the slab is constant (an outgoing plane wave), but on the left the magnitude shows oscillatory character (with incoming and nonzero outgoing waves).

For the wavefunctions plotted in Fig. 4, the complex phase has been chosen such the real part of the wavefunctions attain maximum amplitude within the graphene slab and the imaginary part is minimal. Focusing our attention on the real part, the extrema that occur in the wavefunction between the graphene planes are labeled by "+", "0", or " –" in Fig. 4 in accordance with their projections onto a set of approximate basis functions; these basis functions are pictured in Fig. 5. We denote these as $\xi_{m,p}$ where $m = 1,2,...,n$ denotes the location of a graphene layer within an $n$-layer slab, and $p = L$ or $R$ correspond to the left- or right-hand side of the $m$th layer. We emphasize that these basis functions are *not* used for any quantitative purposes in this work, but rather, we present them simply to provide some definiteness to our discussion. These basis functions are constructed in an approximate manner by taking sums and differences of the eigenstates of Fig. 4. For example, the sum and difference of the image-potential states 4(a) and 4(b) yield the functions on the left- and right-hand ends of the slab, $\xi_{1,L}$ and $\xi_{n,R}$ of Fig. 5 (with $n = 3$ in Fig. 5 whereas $n = 4$ in Fig. 4). Taking further linear combinations, the sum or difference between the states of Figs. 4(c) and 4(f) allows us to localize a function (call it "f1") on the right of plane 2 or the left of plane 3, whereas the sum of 4(c) and 4(e) allows us to form another function (call it "f2") with in-phase contributions between planes 2 and 3 but with zero contributions on the left of plane 2 and on the right of plane 3. Then, shifting function f2 to the right or left by one interplanar separation and subtracting from f1 enables the formation of the basis functions in the interior of the slab, $\xi_{2,L}$ and $\xi_{2,R}$ of Fig. 5. We truncate these inner basis functions at one interplanar spacing on either side of the $m$th plane, and we have, where needed, slightly attenuated the functions at their ends so that they smoothly approach zero there.

Employing the basis functions of Fig. 5, we can form linear combinations of them using coefficients having signs given by the "+", "0", or "–" labels of Fig. 4 to construct the complete states of Fig. 4. For the case of the propagating states 4(c) – 4(h), the linear combinations thus formed are then matched to the oscillatory form in the vacuum using the procedure specified in the following Section (where we also justify our present focus on the real part of the wavefunctions). There are two states of Fig. 4 that require additional comment, namely, (b) and (d). In both cases, the wavefunctions are asymmetric, with a sign change in their coefficient occurring at the midpoint of the slab. It may not be obvious from a first inspection of the wavefunctions whether the coefficients between planes 2 and 3 should be "0,0" (as shown in Fig. 4) or " –,+". However, when we form the states explicitly by summing the basis functions of Fig. 5, we do indeed find that the "0,0" combination is most appropriate for these cases.

Our predicted of reflectivity spectra such as that shown at the top of Fig. 4 are in good agreement with experiment, at least for the first band of transmission resonances (0 – 8 eV) as described in Ref. [9]. For the second band (14 – 22 eV) a similar level of agreement is not apparent [21]. However, at these higher energies, the attenuation of the electron beam (by absorption and/or other decoherence effects) may become more important. As mentioned in Section I, we do not include any imaginary term in the potential that would enable exploration of such effects.

## III. Tight-binding Model

Given the agreement between experiment and the first-principles theory (for the first band), it might seem unnecessary to seek approximate models for describing the reflectivity. However, given the common usage of such models in prior work [1,2,6,18,19], we consider it worthwhile to investigate their applicability for graphene (additionally, simple models may lead to a better qualitative understanding of the phenomenon). The early work of Hibino et al. for the reflectivity of graphene proposed the use of a tight-binding model to understand the spectra [1]. Although their basis functions were localized *on* the graphene planes, and we now know that the correct location is *between* the planes [9], it is nevertheless interesting to consider whether such a model (with appropriate basis functions) can be applied to understanding the spectra. Perhaps the name "tight-binding" is inappropriate for the model we will construct, since the basis functions are not formed from atomic orbitals of the atoms. Rather, we will construct basis functions from the interlayer states of the material, as already described above in connection with Fig. 5. Such functions, by their nature, tend to be strongly pinned to the location of the atoms, and for this reason they can be treated in a tight-binding method. While these basis functions display a certain amount of nonorthogonality, the method for handling that is well known [22] and will be employed in our solution.

One initial point to be considered in this procedure is that tight-binding models themselves generally describe *bound states*, whereas the states at the reflectivity minima that we are interested in are not bound, but rather, they are *scattering states* of the system. To understand how a tight-binding model can be applied in this case, let us first recall the situation for a uniform potential well as described in textbook treatments: the minima in reflectivity that occur in the scattering states of a finite potential well, i.e. the *transmission resonances*, have identical energy as the bound states that form in the infinite potential well problem [23]. The reason that these energies are identical is clear from a consideration of the respective boundary conditions in the two situations: For the infinite well, the wavefunction is of course zero at the boundaries, thus having the form $\sin(kx)$ with $k = \pi m/L$, where $m$ is a positive integer and with the boundaries of the well at $z = 0$ and $z = L$. The corresponding energy relative to the bottom of the well is $E = \hbar^2 k^2 / 2m_e$, with $m_e$ being the electron mass. Within the potential well we have the linearly independent solution of $\cos(kx)$, but those states do not satisfy the boundary condition for the infinite well. Now let us construct the scattering states of the finite well. For the $\sin(kx)$ solutions, we extend these into the vacuum simply by matching their derivatives to the states outside the well, yielding $k \sin(k'x)/k'$ outside the well with $k' = \sqrt{2m_e(E-V_0)}/\hbar$ where $V_0 > 0$ is the depth of the well. For the $\cos(kx)$ solutions they are extended by matching the value of the wavefunctions at the boundary, yielding $\cos(k'x)$ outside the well. Then by taking linear combinations of the states outside the well thus formed, including a scaling of the cosine solutions by $k/k'$, we achieve a full description of the scattering states. The crucial point is that the sine and cosine solutions for the finite well have the *same* phase shifts at the boundaries (zero phase shift in this case) so that, as in Eq. (3), the reflectivity is zero.

We can apply a similar argument to forming tight-binding type solutions for describing the states of our multilayer graphene. For *n* layers of graphene, we will form solutions using 2*n* basis states. A convenient basis, consisting of states that are localized on the left- and right-hand sides of each graphene plane has already been introduced in Fig. 5. States constructed from these functions must be matched to sine or cosine type propagating solutions extending out into the vacuum. This matching can be accomplished, e.g., at the position where the image-potential state has a maximum. Referring to the real part of the wavefunction in curve (c) of Fig. 4, this position on the left-hand side of the slab is indicated by an arrow; the wavefunction to the left of that would have a maximum at this matching point. Importantly, the imaginary part of the wavefunction at this point has a zero, i.e. it has the *same phase shift* as the real part. Thus, the solutions formed by linear combinations of the basis functions serve as an approximation for the real part of the wavefunction, and we know from the full first-principles computation that an appropriate imaginary part for the wavefunction also exists such that the complete scattering state can be constructed.

Using states consisting of linear combinations of the basis functions $\xi_{m,p}$ of Fig. 5 within a tight-binding type framework, we parameterize the Hamiltonian matrix **H** and the orthogonality matrix **R**, with $H_{m,p,m',p'} \equiv \langle \xi_{m,p} | \hat{H} | \xi_{m',p'} \rangle$ and $R_{m,p,m',p'} \equiv \langle \xi_{m,p} | \xi_{m',p'} \rangle$ where $\hat{H}$ is the total Hamiltonian for the system. Numerical values for these matrix elements are chosen to provide a fit to the spectrum of transmission resonances. With these values, we then solve for the eigenvectors **ψ** and eigenvalues $E$, using $\mathbf{H\psi} = E\,\mathbf{R\psi}$ so that $\mathbf{R}^{-1}\mathbf{H\psi} = E\,\mathbf{\psi}$. We therefore need only evaluate the eigenvectors and eigenvalues of the (non-Hermitian) matrix $\mathbf{R}^{-1}\mathbf{H}$.

The energy locations of the transmission resonances from the first-principle computations are summarized in Fig. 6, and we choose our tight-binding parameters to match those energies. There are two bands that are separated by an "on-site" matrix element. The central energy of the bands, $\langle \chi_{m,p} | \hat{H} | \chi_{m,p} \rangle$ with $(m,p) \neq (1, L)$ and $(n, R)$, is taken to be 10.4 eV. For $(m,p) = (1, L)$ or $(n, R)$, this energy is $-0.4$ eV, corresponding to the image-potential states at the ends of the slab. The hopping matrix element, $\langle \chi_{m,p} | \hat{H} | \chi_{m,p'} \rangle$ with $p \neq p'$, determines the width of each band, and is taken to be $-3.0$ eV for $m \neq 1$ and $n$. If $m = 1$ or $n$, then this element is $-0.4$ eV for $n = 1$, and $-0.8$ eV for $n \neq 1$. The other important matrix element, $\langle \chi_{m,R} | \hat{H} | \chi_{m+1,L} \rangle$ with $m \neq n$, can be viewed as an "on-site" term since it couples states that overlap in the same region of space (i.e. between graphene planes). This matrix element determines the separation of the two bands, and it is taken to be $-7.5$ eV. All other matrix elements of **H** are taken to be zero. Concerning the orthogonality matrix, the only term that we find to be important is $\langle \chi_{m,p} | \chi_{m,p'} \rangle$ with $p \neq p'$ and $m \neq 1$ and $n$, which we take to be 0.04 (this value affects the relative widths of the upper and lower bands).

With these seven parameters, the tight-binding model produces energies that match quite well to the first-principles results, as seen in Fig. 6. Additionally, inspection of the eigenvectors from the model also reveals good agreement with the labeling of the states in Fig. 4. We conclude that this sort of tight-binding model provides a good, approximate description of the transmission resonances in a graphene slab.

In principle, this tight-binding model can be generalized to include the presence of a substrate on one side of the graphene slab. That change would entail a new set of parameters to describe the matrix elements on one side of the slab. However, experimentally, it is primarily only the first band of interlayer states (0 – 8 eV) that is observed, and hence we restrict our attention to only that energy range. In this case, examining the states of Fig. 4, we see that in the spaces between the graphene planes, there are always symmetric combinations of the basis functions formed, i.e. (+,+), (0,0), or (–,–). Hence, it is convenient to change our basis such that it is formed from the appropriate combinations of our basis states, i.e. $\xi_{1,R} + \xi_{2,L}$ between the 1st and 2nd planes, etc., and where we need not include the end functions $\xi_{1,L}$ and $\xi_{n,R}$ in this reduced description. In this way, for a free-standing slab we end up with $n-1$ new basis states for $n$-layer graphene. With a substrate there is a new space that is formed between the bottommost graphene plane and the substrate, and in this space an additional interlayer state (basis function) can form. This new state will have, in general, a different energy than the others, and in the reflectivity spectrum a new minimum will be formed. Results of this sort of tight-binding model, with $n$ states for $n$ layers of graphene (with a possible graphene-like layer terminating the substrate not included in that count), are presented elsewhere [9,10].

## IV. Discussion

For illustrative purposes, we consider another possible model for describing the reflectivity, namely, by treating the potential of Fig. 1(c) in a 1D model and integrating the 1D Schrödinger equation using this potential to solve for the reflectivity vs. energy. Figure 7(a) shows the result, which deviates significantly from the first-principles spectrum of Fig. 4. Bragg peaks (near 10 and 35 eV) are seen in the 1D results, in agreement with experiment [21], but an expected Bragg peak near 22 eV is *not* seen in the 1D result. This latter Bragg peak arises from the AB Bernal stacking of graphite, which induces a component of the potential at the $c = 0.671$ nm lattice spacing (i.e. twice the interplanar distance), something that is not present in the plane-averaged potential of Fig. 1(c) which has $c/2$ periodicity. Aside from the Bragg peaks, Fig. 7(a) reveals reflectivity zeroes at about 0.3 and 6.0 eV (1st band) and 17 and 27 eV (2nd band), qualitatively in agreement with the first-principles results although the zeroes of the 2nd band are too high in energy. Additionally, the expected zero at about 3.7 eV (1st band) is only seen as an inflection point, and one at 21 eV (2nd band) only appears as a weak minimum. More significantly, between these local minima, the transmission rises to values of only about $2\times10^{-2}$ or $2\times10^{-4}$ (1st band) or $2\times10^{-5}$ (2nd band), which is in contrast to the first-principles results where the transmission rises to values of order unity between the local minima.

These differences between the 1D model and the first-principles computations arise, we believe, because the 3D nature of the potential *is* important.[2] We have tried modifying the 1D potential in an effort to mimic the 3D results, e.g. by adding a 10-eV-deep potential well with width of 0.2 nm on each of the four locations corresponding to the graphene planes. The situation is then slightly improved, as shown in Fig. 7(b). The transmission now rises to values of $10^{-3} – 10^{-1}$ in the regions between the reflectivity minima. Note, however, that in this case that the lowest reflectivity minimum moves down in energy [since the average potential is now deeper than for Fig. 7(a)], to about 0.1 eV, which is almost off the plot. If we arbitrarily scale the potential of Fig. 2(c) by a factor of 0.98, and then add the 10-eV-deep wells at the positions of the graphene planes, the spectrum of Fig. 7(c) is produced. The reflectivity minima have now shifted up slightly in energy. In general, we find the reflectivity obtained by 1D integration to be rather sensitive to the precise form of the potential, and for this reason we have not further pursued this type of model.

---

[2] The extrema in the 3D wavefunctions of Fig. 4 that are located at or near the midpoints between the graphene planes have nearly no $x$ or $y$ dependence, whereas the extrema located close to the atomic cores *do* depend significantly on $x$ and $y$. These latter features are not well represented in a 1D model.

Another model that has been commonly used to describe reflectivity spectra from a range of thin films is the phase accumulation (PA) model [18,19]. This model essentially treats the potential within the film as a constant, and quantum well resonances are formed in the film. Good success has been achieved in applications of the model to thin metal films on surfaces, where the concept of a constant potential in the metallic film (i.e. equal to the *inner potential*) seems reasonable [18,19]. For graphene, however, it is less clear whether or not the highly varying potential throughout the material permits the PA model to be applicable. Two prior applications of this type of model to multilayer graphene have been made, although very different parameter values were employed in the two cases [2,6]. Another factor that merits consideration is whether the resonances predicted by the PA model correspond to reflectivity *minima* or *maxima* (since both cases have been used in prior applications [1,2,6,18,19]); we will assume the former here, i.e. transmission resonances, based on both the discussions of Hibino et al. [1] and of Section III.

In the PA model, the energy of transmission resonances is assumed to be given by solutions to the equation

$$2k(E)nt + \Phi(E) = 2m\pi \qquad (4)$$

where $k(E)$ is the perpendicular wavevector in the film, $n$ is the number of atomic layers, $t$ is the average layer spacing, $\Phi(E)$ is the energy dependent phase shift accumulated upon reflections at the boundaries of the film (i.e. by an electron bouncing back and forth within the film), and $m$ is an integer labeling the particular state (note that our usage of $n$ and $m$ is interchanged compared to prior authors, such that it is more consistent with our tight-binding description of Section III). For a given value of $n$, solutions to this equation for various $m$ have provided a reliable means of measuring $k(E)$ for thin metallic films [18,19]. The influence of the $\Phi(E)$ term can be eliminated in these solutions by examining the *difference* in $m$ values for resonances that occur at the same energy but for different (known) $n$ values, thus providing a reliable means of employing the PA model (although, nevertheless, some knowledge of the $\Phi(E)$ term does exist for specific surfaces and interfaces) [19].

It is important to realize that for the prior work on thin metallic films, many of the quantum well resonances formed in the films lie well below the vacuum level. It is only higher-lying states that are observed by LEEM reflectivity. This spectrum of states is apparent if we take the $k$ values from Eq. (4) and obtain an energy (relative to the vacuum level) from a free electron model of $(\hbar^2 k^2 / 2m) - V_0$ where $V_0 > 0$ is the inner potential. For a typical inner potential of $\approx 14$ eV for a metal, it is only solutions of Eq. (4) with high $m$ values that have positive energies and hence can be matched to the observed LEEM reflectivity extrema.

Here, we consider the case of multilayer graphene. Specifically, examining the states of Fig. 4, we see that they can be decomposed according to the number of $\pi$ phase changes in the wavefunctions. For example, consider Fig. 4(c) starting from the zero in the real part of the wavefunction that occurs immediately to the right of the arrow, at $z = -0.657$ nm. From that point, over to the corresponding zero at $z = +0.657$ nm, there are 9 half-oscillations of the wavefunction, i.e. $m = 9$ in Eq. (4) for this situation of $n = 4$ layers of graphene. The states in Figs. 4(d) and 4(e) have $m$ values of 10 and 11, respectively. The locations of the zeroes on either side of the slab occur at a distance of 0.154 nm from the nearest graphene plane, which is fairly close to half of the interplanar separation of 0.3355/2 = 0.168 nm. Hence, to a reasonably good approximation, we can take the value of $\Phi(E)$ in Eq. (4) to be zero, and therefore we have simply $k = m\pi / nt$ where $t = 0.3355$ nm. Using the free electron model, and assuming some inner potential, we then obtain the energies of the states.

Figure 8 shows result of these PA model computations employing an inner potential of 18 eV [6]. For the 1st band, we obtain generally the correct shape of the band, although its spread (dispersion) is too large compared to the first-principles results of Fig. 6. However, if we follow the same procedure for the 2nd band the results are much worse, as seen in Fig. 8. The reason for this discrepancy can be traced to the location of the zeroes in the wavefunctions for the states in Figs. 4(f) – 4(h). These are much closer to the nearest graphene planes than for the states 4(c) – 4(e). In terms of Eq. (4), to accommodate this property we would have to assume nonzero $\Phi(E)$ values. However, we find that the values needed in order to produce an arrangement of states in this 2nd band that resembles those of Fig. 6 turn out to depend not only on the energy, but also on the $n$ value.[3] The PA model thus fails to describe the 2nd band. Similarly, the image-potential states of the first-principles results have no analog in the PA model, and the predictions of Eq. (4) using $n = 1$ have no correspondence with the first-principles results.

Thus we conclude that the PA model provides a reasonable, qualitative description of the data (for the first band), and in this respect we agree with the analysis of Sutter et al. [6]. However, our detailed interpretation utilizing the PA model differs from theirs in terms of the $m$ values used to label the states. For example, for 3-layer graphene they label the reflectivity minima by 15/2=7.5 and 17/2=8.5, corresponding to our values of 7 and 8, respectively. For 4-layer graphene, they employ 21/2=10.5, 23/3=11.5 and 25/2=12.5, deviating even more from our values of 9, 10, and 11, respectively. Similarly, their higher-lying states have $m$ values that deviate substantially from ours. In our case there is no uncertainty in the *m* values, since they are obtained directly from the first-principles wavefunctions. Hence, we feel that the fits of Sutter et al. are not sufficiently constrained in terms of *m* values, hence leading to their apparently good fit between the PA model (employing free electron energies) and the experimental data [6].

---

[3] To achieve a reasonably good spectrum of states for the 2nd band based on Eq. (4), we modify that equation so that rather than having a distance of $nt$ in the first term we use the actual extent of the wavefunction between the two zeroes, $(n - 2f)t$ with $f = 0.5 - (|\Delta z|/t)$ where $\Delta z$ is the location of the zero relative to the nearest graphene plane. Phenomenologically, we find that a value of $f = 0.15$ produces a good overall spectrum, although the location of the band center turns out to be $\approx 6$ eV too high in energy and its spread is also too large.

Locatelli et al. also applied the PA model to understand the graphene reflectivity spectrum, again employing Eq. (4). In contrast to Sutter et al., however, they implicitly assumed a much smaller effective average potential in the material. The entire spectrum of states predicted from the model was assumed to lie *above* the vacuum level, i.e. including all states even with low *m* values of 1,2,3,…. We feel that this interpretation is, on a purely qualitative basis, consistent with our first-principles (and tight-binding) descriptions of the transmission resonances in that they are a special set of interlayer states with no corresponding states that lie below the vacuum level. An interpretation of the wavefunctions of Fig. 4 using values of $m = 1,2,3,...$ clearly applies only to the *envelope functions* of the combined basis functions discussed at the end of Section III. That is, in terms of the combined basis functions (and neglecting the end states), Fig. 4(c) would have weights of (+,+,+), Fig. 4(b) would have (−,0,+), and Fig. 4(c) would have (−,,+,−), with these weights interpreted as envelope function values at the midpoints between pairs of neighboring graphene planes. This approach is much different from that employed in previous applications of the PA model [6,18,19] but again, we feel that it is in qualitative accord with our first-principles results.

To achieve quantitative agreement of this type of PA model with the first-principles results, it is necessary utilize nonzero values for $\Phi(E)$. To illustrate this, we assume initially a constant value of $\Phi(E) = \pi$. Figure 9 displays that spectrum of states from Eq. (4), with the energies in this case given by simply $\hbar^2 k^2 / 2m_e$ (i.e. with no inner potential shift). Qualitative agreement with the first-principles results (Fig. 6) for the 1st band is obtained, although the width of the PA-predicted band is too small. We do not attempt modeling of the 2nd band, i.e. utilizing higher *m*-values in Eq. (4), since the resulting energies are much too low and to correct them would involve a further assumption of an energy shift of the 2nd band, something that goes well beyond the range of applicability of the PA model. To justify our assumed value of $\Phi(E) = \pi$, we note that the eigenvectors for a tight-binding model of a 1D chain do indeed lead to phase shifts close to that. The tight-binding model in this case reduces to a matrix with values $\varepsilon_0$ on the diagonal and $-t$ on the upper and lower off-diagonals, but with zero coupling between the ends. Examining the eigenvectors of that we find that they are independent of the values of $\varepsilon_0$ and *t*, so long as $t \neq 0$. Matching their values to a sine or cosine form relative to the center of the chain leads to a phase shift at the ends of the chain, with $\Phi(E)$ being four times that value. In this way we find $\Phi(E)$ values of about $1.1\pi$ at an energy near the center of the band, and ranging from about $0.3\pi$ to $1.9\pi$ at the bottom and top of the band, respectively.

We conclude that the PA model *is* applicable to the reflectivity of multilayer graphene, and it can be applied in two quite different ways – one proposed by Sutter et al. [6] and the other by Locatelli et al. [2]. Figures 8 and 9 illustrate the results, employing free electron dispersion of the states. Of course, it is better to utilize the *difference method* as previously developed for thin metallic films [18,19], in which case no assumption of the $k(E)$ values or $\Phi(E)$ is needed. This type of analysis was indeed performed in Ref. [2], although we disagree with the *m* values assigned to the reflectivity minima in that work.[4] The difference method was not employed in Ref. [6], but in any case we again disagree with the *m* values employed there. This initial assignment of *m* values (or, more precisely, $\Delta m$ values) in the difference method is one of the challenges encountered when utilizing the PA model. In our first-principles analysis, the *m* values arise directly from the computed wavefunctions so that they can be assigned with certainty.

**V. Summary**

In summary, we have developed a method for obtaining the low-energy reflectivity spectra of materials based on a first-principles evaluation of their electronic structure. The method is most useful for 2D materials, in which interlayer states form and contribute to the spectrum of states just above the vacuum level [12]. Good agreement between experiment and theory is found for multilayer graphene. The first-principles results are interpreted within the framework of a tight-binding model. Introducing localized states on the right- and left-hand sides of each graphene plane, we obtain 2*n* states for an *n*-layer graphene slab. Solving for the eigenstates of the system, the two lowest energy states are composed primarily of image-potential states on either end of the graphene slab. The remaining $2(n-1)$ states are divided into two bands, one occurring in the range 0 – 8 eV and the second at 14 – 22 eV. It is demonstrated that a tight-binding description of these states is in reasonable agreement with the first-principles results.

**Acknowledgements**

We are grateful to N. Srivastava and I. Vlassiouk for useful discussions, and we thank P. Mende S. de la Barrera for their careful reading of the manuscript. This work was supported by the National Science Foundation under grant DMR-1205275, and by the Office of Naval Research MURI program under grant N00014-11-1-0678.

---

[4] The assignment of the reflectivity curves of Ref. [2] to various thicknesses of multilayer graphene has been revised from the original work; see Ref. [9] for a discussion.

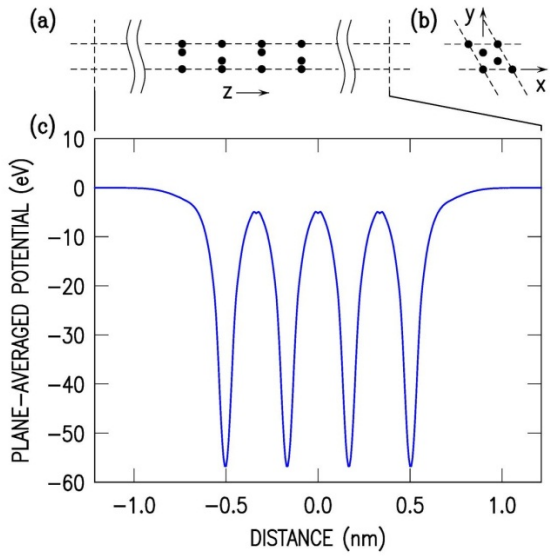

FIG 1. (Color on-line). (a) and (b) Atomic positions (black dots) in the simulation slab used in the first-principles computations, for 4 layers of graphene. (c) Computed potentials, averaged over the *xy* plane.

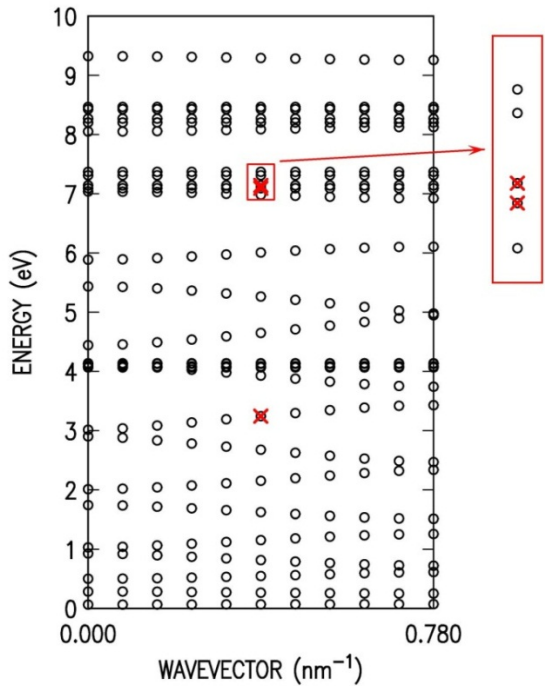

FIG 2. (Color on-line). Computed energies of electronic states for a 4-layer graphene slab with potential shown in Fig. 1(c), for a range of energies 0 – 10 eV above the vacuum level. Three energies are highlighted (two of them shown more clearly in the inset), with wavefunctions for these states shown in Fig. 3.

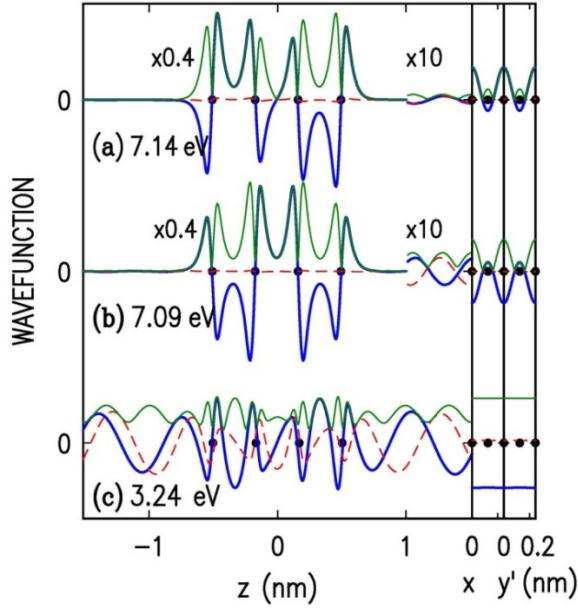

FIG 3. (Color on-line). Wavefunctions for the states whose energy locations are highlighted in the band structure of Fig. 2, plotted with respect to the axes defined in Fig. 1, with $y' = y - (x/\sqrt{3})$. The plots vs. $z$ are evaluated at the values $x = y = 0$, whereas the plots vs. $x$ and $y'$ are evaluated at a $z$-value that is midway between the left-most and the neighboring graphene planes. The real part of the wavefunctions are shown by a thick (blue) solid line and the imaginary part by a thin (red) dashed line, with the magnitude shown by a thin (green) solid line. For plots (a) and (b), insets on the far right-hand side of the $z$-axis show expanded views of the wavefunctions. In all plots, solid black dots indicate positions of carbon atoms.

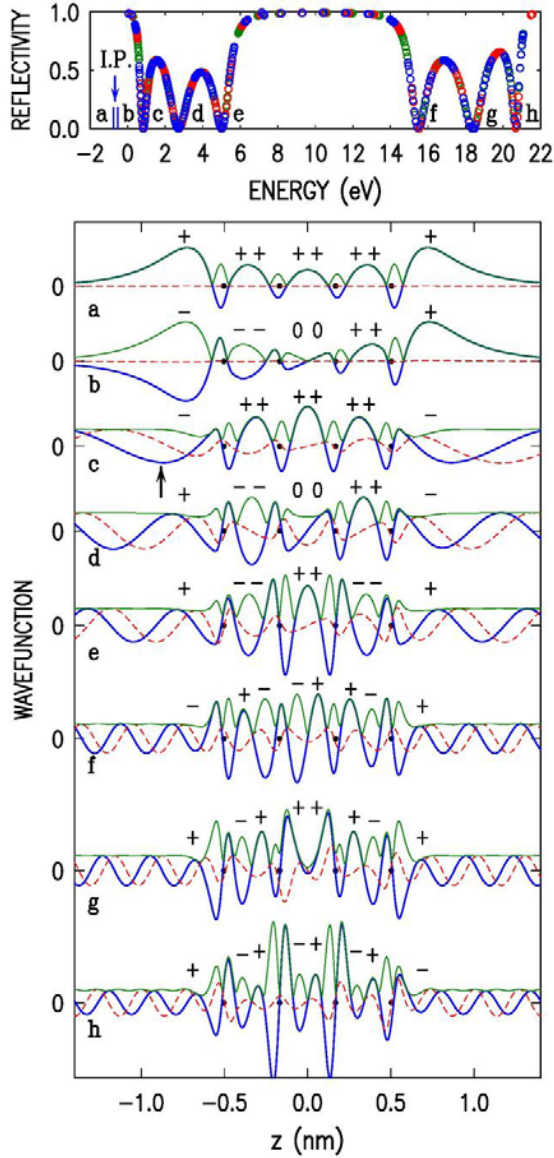

FIG 4. (Color on-line). The upper panel shows the computed reflectivity for a free-standing slab of 4-layer graphene. A series of computation are performed with different vacuum widths; differently shaded (colored) data points are used for plotting the results for each width. Energy positions (a,b) for the two image potential (I.P.) states are indicated, and similarly (c,d,e) label the zeroes of the reflectivity in the first band, and (f,g,h) label the zeroes in the second band. Corresponding wavefunctions are shown in the lower panel, with the real part of the wavefunctions indicated by a thick (blue) solid line, the imaginary part by a thin (red) dashed line, and the magnitude by a thin (green) solid line. The labeling of extrema of the real part of the wavefunctions by "+", "0", or "−" is as described in the text. The arrow for curve c indicates a location for connecting a localized state to a scattering state, as also discussed in the text. Solid black dots indicate the positions of carbon atoms.

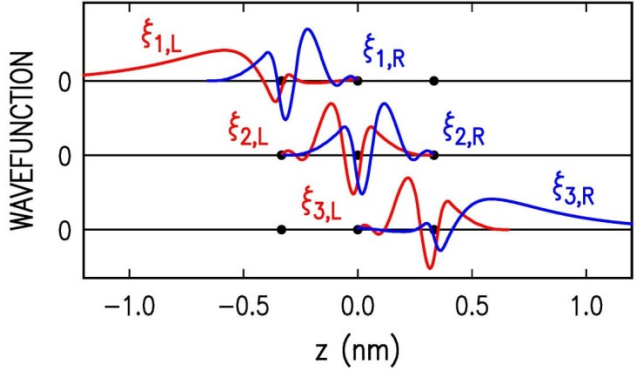

FIG 5. (Color on-line). A set of approximate, localized basis functions for constructing the complete wavefunctions of the transmission resonances, shown for a 3-layer graphene slab. Basis functions are denoted by $\xi_{m,p}$ with $m$ referring to the graphene layer number and $p = L$ or $R$ referring to the left- or right-hand sides of the layer, respectively. Different basis functions are employed at the ends of the graphene slab, i.e. $(m, p) = (1, L)$ or $(n, R)$, compared to the interior of the slab. Solid black dots indicate the positions of carbon atoms.

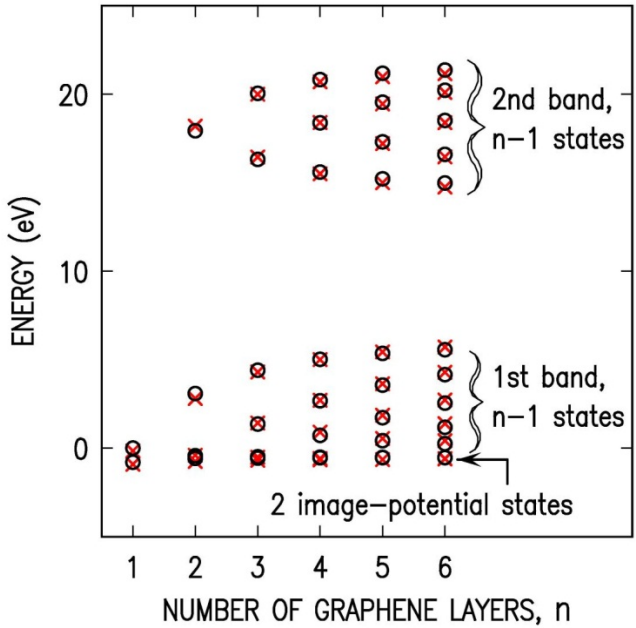

FIG 6. (Color on-line). Energy of reflectivity minima from first-principles computations (x-marks) and predictions of tight-binding model (circles), as a function of the number of graphene layers in a free-standing slab, $n$. The two lowest energy states are derived from the image potential at the ends of the graphene slab, with $n-1$ states in each of the 1st and 2nd bands. All energies are measured relative to the vacuum level.

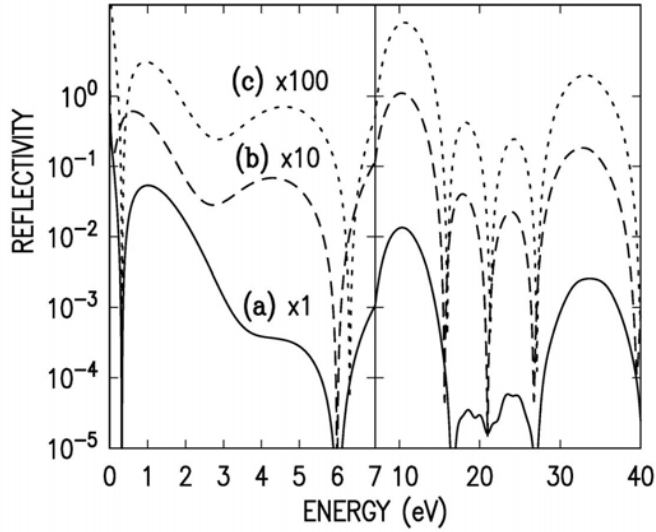

FIG 7. Computed reflectivity, based on integration of a 1D Schrödinger equation. Curve (a) uses the plane-averaged potential from Fig. 1(c), with curves (b) and (c) including modifications to that potential as described in the text. Note the change in energy scale at 7 eV.

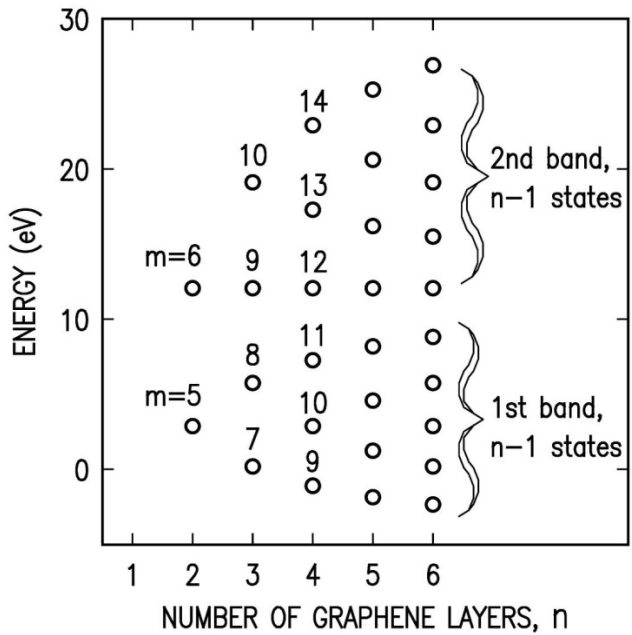

FIG 8. Energy of reflectivity minima from a phase accumulation model, utilizing free electron energies relative to a band minimum at 18 eV below the vacuum level, as a function of the number of graphene layers in a free-standing slab. A boundary phase of $\Phi(E) = 0$ is assumed (with more accurate values discussed in the text). Some points are labeled by their $m$ values, as defined in the text.

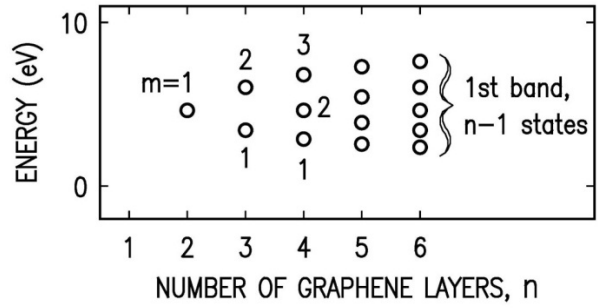

FIG 9. Energy of reflectivity minima from a phase accumulation model, utilizing free electron energies relative to the vacuum level, as a function of the number of graphene layers in a free-standing slab. A boundary phase of $\Phi(E) = \pi$ is assumed (with more accurate values discussed in the text). Some points are labeled by their *m* values, as defined in the text.